\documentclass[aps,prl,twocolumn]{revtex4-1}
\pdfoutput=1
\usepackage{graphicx}
\usepackage{bm} 

\newcommand{\imgsize}{7.4cm}   

\begin{document}

\title{Thermal $\Upsilon(1s)$ and $\chi_{b1}$ suppression in $\bf \sqrt{s_{NN}}=2.76$ 
TeV Pb-Pb collisions at the LHC }

\author{Michael Strickland}
\affiliation{Department of Physics, Gettysburg College, Gettysburg, Pennsylvania 17325, USA}
\affiliation{Frankfurt Institute for Advanced Studies, D-60438 Frankfurt am Main, Germany}

\date{\today}

\begin{abstract}
I compute the thermal suppression of the  $\Upsilon(1s)$ and $\chi_{b1}$ states in 
$\sqrt{s_{NN}}=2.76$ TeV Pb-Pb collisions.  Using the suppression of each of these states
I estimate the total $R_{AA}$ for the $\Upsilon(1s)$ state as a function of centrality, rapidity,
and transverse momentum.  I find less suppression of the $\chi_{b1}$ state than would be
traditionally assumed; however, my final results for the total $\Upsilon(1s)$ suppression are in 
good agreement with recent preliminary CMS data.
\end{abstract}


\maketitle


The behavior of nuclear matter at extreme temperatures is now being studied with the 
highest collision energies ever achieved using the Relativistic Heavy Ion Collider 
at Brookhaven National Laboratory and the Large Hadron Collider at CERN.  For RHIC 
$\sqrt{s_{NN}} = 200$ GeV Au-Au collisions, initial central temperatures of $T_0 \sim 350$ MeV
were generated.  For current LHC $\sqrt{s_{NN}} = 2.76$ TeV collisions one obtains 
$T_0 \sim 500-600$ MeV \cite{Schenke:2011tv} and for upcoming full energy runs with 
$\sqrt{s_{NN}} = 5.5$ TeV one expects $T_0 \sim 700-800$ MeV.  At such high temperatures, one
expects to generate a quark-gluon plasma (QGP) in which the formation of quark bound
states is suppressed in favor of a deconfined plasma of quarks
and gluons.

Suppression of quark bound states follows from the fact that in the 
QGP one has Debye screening of color charge \cite{Shuryak:1980tp,%
*Matsui:1986dk,*Karsch:1987pv}.  Heavy quarkonium has received the most
theoretical attention, since heavy quark states are dominated by short distance 
physics and can be treated using heavy quark effective theory.  Based on such effective theories 
of QCD, non-relativistic quarkonium states can be reliably
described. Their binding energies are much smaller than the quark mass $m_Q\gg\Lambda_{\rm
  QCD}$ ($Q=c,b$), and their sizes are much larger than $1/m_Q$. At zero
temperature, since the velocity of the quarks in the bound state is
small ($v\ll c$), quarkonium can be understood in terms of
non-relativistic potential models such as the Cornell potential which can
be derived directly from QCD using effective field theory 
\cite{Eichten:1979ms,*Lucha:1991vn,* Brambilla:2004jw}.

Using such non-relativistic potential models,
studies of quarkonium spectral functions and meson current correlators
have been performed
\cite{Mocsy:2004bv,*Wong:2004zr,*Cabrera:2006wh,*Alberico:2007rg,*Mocsy:2007yj}.
The results have been compared to first-principles lattice QCD calculations
\cite{Umeda:2002vr,*Asakawa:2003re,%
*Datta:2003ww,*Aarts:2006nr,*Hatsuda:2006zz,*Jakovac:2006sf,%
*Aarts:2010ek} which rely on the maximum entropy method 
\cite{Nakahara:1999vy,*Asakawa:2002xj}.
In recent years, however, there has been an important theoretical development,
namely the first-principles calculation of the thermal widths of heavy quarkonium states
which emerge from imaginary-valued contributions to the
heavy quark potential.  The first calculation of the leading-order
perturbative imaginary part of the potential due to gluonic Landau damping was performed by Laine et 
al.~\cite{Laine:2006ns,*Laine:2007gj}.
Since then an additional imaginary-valued contribution to the potential coming from singlet to
octet transitions has also been identified~\cite{Brambilla:2008cx}.
The consequences of such imaginary parts on heavy quarkonium spectral functions 
\cite{Miao:2010tk}, perturbative thermal widths \cite{Laine:2006ns,*Brambilla:2010vq}, and 
quarkonium suppression in a T-matrix approach \cite{Riek:2010py} 
have recently been studied; however, these studies were restricted to the case of an isotropic
thermal plasma, which is only the case if one assumes ideal hydrodynamical evolution.

The calculation of the heavy quark potential has since been extended to the case of a plasma
with finite momentum-space anisotropy for both the real \cite{Dumitru:2007hy,*Dumitru:2009ni}
and imaginary \cite{Burnier:2009yu,*Dumitru:2009fy,*Philipsen:2009wg} parts.  Additionally, the
impact of the imaginary part of the potential on the thermal widths of the states in both isotropic
and anisotropic
plasmas was recently studied~\cite{Margotta:2011ta}.  The consideration of momentum-space
anisotropic plasmas is necessary since, for any finite shear viscosity, the quark-gluon plasma possesses
local momentum-space anisotropies
\cite{Israel:1976tn,*Israel:1979wp,*Baym:1984np,*Muronga:2003ta,%
*Florkowski:2010cf,*Ryblewski:2010bs,*Ryblewski:2011aq,*Martinez:2010sc,Martinez:2010sd}.
Depending on the magnitude of the shear viscosity, these momentum-space anisotropies can persist
for a long time and can be quite large, particularly at early times or near the edges of
the plasma.  

This has motivated the development of a new dynamical formalism called ``anisotropic
hydrodynamics'' ({\sc aHydro}) which extends traditional viscous hydrodynamical treatments 
to cases in which the local momentum-space anisotropy of the plasma can be large
\cite{Florkowski:2010cf,*Martinez:2010sc,*Ryblewski:2010bs,*Ryblewski:2011aq,Martinez:2010sd}.
In this Letter I present first results of the combination of the
{\sc aHydro} temporal evolution of Ref.~\cite{Martinez:2010sd} together with results obtained in 
Ref.~\cite{Margotta:2011ta} for the real and imaginary parts of the binding energy.
Using this framework I compute the suppression of the $\Upsilon(1s)$ and $\chi_{b1}$
mesons as a function of centrality, rapidity, and transverse momentum.


\vspace{1mm}

\noindent{\sc Model potential:}
The phase-space distribution of gluons in the local rest frame is assumed to be given by 
$f({\bf x},{\bf p}) = f_{\rm iso}\left(\sqrt{{\bf p}^2+ \xi({\bf p}\cdot{\bf n})^2 }  / 
p_{\rm hard} \right)$ where $p_{\rm hard}$ is a scale which specifies the typical 
momentum of the particles and can be identified with the temperature
in an isotropic plasma ($\xi=0$) \cite{Romatschke:2003ms}.  In general, the
parameter $\xi$ measures the degree of anisotropy of the plasma via $\xi = \frac{1}{2} \langle 
{\bf p}_\perp^2\rangle/\langle p_z^2\rangle -1$ where $p_z$ and 
${\bf p}_\perp $ are the partonic longitudinal and transverse momenta in the local
rest frame, respectively.  

The perturbative heavy quark potential as function of $\xi$ has been evaluated previously
and has both real \cite{Dumitru:2007hy} and imaginary contributions \cite{Dumitru:2009fy}.  
For $N_c=3$ and $N_f=2$ the real part of the resulting potential can be well-approximated by 
$\Re[V_{\rm pert}] = - \alpha \exp(-\mu r)/r$ with
\begin{equation}
\left(\frac{\mu}{m_D}\right)^{-4} =  
1 + \xi\left(1 + \frac{\sqrt{2}(1+\xi)^2\left(\cos(2\theta) - 1 \right)}{(2+ \xi)^{5/2}} \right) \, ,
\label{eq:muparam}
\end{equation}
where $\alpha = 4\alpha_s/3$, $m_D^2 = (1.4)^2 16 \pi \alpha_s  \, p_{\rm hard}^2/3$ is the isotropic
Debye mass, and $\theta$ is the angle with respect to the beamline.  
The factor of $(1.4)^2$ accounts for higher-order corrections to the isotropic Debye 
mass \cite{Kaczmarek:2004gv}.

However, for describing finite-temperature states which can have large radii compared to
$\Lambda_{\rm QCD}^{-1}$, one must supplement the perturbative short range contribution above
by a long range contribution.  Following previous work \cite{Dumitru:2009ni}, I generalize
the Karsch-Mehr-Satz potential \cite{Karsch:1987pv} by including the anisotropic mass scale $\mu$
given in Eq.~(\ref{eq:muparam}) in place of the isotropic Debye mass and adding the entropy
contribution necessary to obtain the internal energy of the states.  Such a construction agrees well 
with lattice measurements of the real part of the heavy quark potential \cite{Kaczmarek:2004gv}.  The 
resulting model for the real part of the heavy quark potential is 
\begin{eqnarray} 
\label{eq:repot}
\Re[V] &=& -\frac{\alpha}{r} \left(1+\mu \, r\right) \exp\left( -\mu
\, r  \right) + \frac{2\sigma}{\mu}\left[1-\exp\left( -\mu
\, r  \right)\right] \nonumber \\
&& \hspace{2cm} - \sigma \,r\, \exp(-\mu\,r)- \frac{0.8 \, \sigma}{m_Q^2\, r} \, ,
\end{eqnarray}
where the last term is a temperature- and spin-independent quark mass correction 
\cite{Bali:1997am} and $\sigma = 0.223$ GeV is the string tension.  Here I ignore
the effect of the running of $\alpha_s$ and fix $\alpha = 0.385$ to match zero temperature
binding energy data for heavy quark states \cite{Dumitru:2009ni}.

For the imaginary part of the model potential I use the imaginary part of the perturbative heavy quark
potential which has been calculated to linear order in $\xi$ 
\begin{equation} 
\Im[V_{\rm pert}] = -\alpha p_{\rm hard} \biggl\{ \phi(\hat{r}) - \xi \left[\psi_1(\hat{r},
\theta)+\psi_2(\hat{r}, \theta)\right]\biggr\} ,
\label{eq:impot}
\end{equation}
where $\hat{r}=m_D r$ and $\phi$, $\psi_1$, and $\psi_2$ are defined in Ref.~\cite{Dumitru:2009fy}.

The full model potential is given by $V = \Re[V] + i \Im[V]$ and can be used in the Schr\"odinger
equation.  To solve the resulting Schr\"odinger equation I 
use the finite difference time domain method~\cite{Strickland:2009ft} extended
to the case of a complex-valued potential~\cite{Margotta:2011ta}.  
For the temperature and anisotropy
dependence of the resulting real and imaginary parts of  the binding energies for the $\Upsilon(1s)$ 
and $\chi_{b1}$, I refer the reader to Ref.~\cite{Margotta:2011ta}.
For a point of reference, in an isotropic plasma the medium-induced width of the $\Upsilon(1s)$ 
is approximately $\Im[E_{\rm bind}] \sim 0.211\,T$.


\vspace{1mm}

\noindent{\sc Initial conditions and dynamics:}
Solution of the Schr\"odinger equation gives the real and imaginary parts of 
the binding energy of the states.  The imaginary part defines the instantaneous width of the state
$\Im[E_{\rm bind}(p_{\rm hard},\xi)] \equiv -\Gamma_T(p_{\rm hard},\xi)/2$.
However, one must account for the complete disassociation of the states when 
$\Re[E_{\rm bind}]<0$.  I implement this by assigning a large
width of 10 GeV $\sim$ (0.02 fm/c)$^{-1}$ to states when $\Re[E_{\rm bind}]<0$.  The final results are
insensitive to the value of this rate, as long as it is taken to be greater than 0.5 GeV.  Given the binding 
energy data, I evolve the system
using the non-boost invariant {\sc aHydro} equations of Ref.~\cite{Martinez:2010sd}.  
Using the output I can compute the temporal 
evolution of the thermal width which gives $\Gamma_T(\tau)$. 

The resulting width $\Gamma_T(\tau)$ implicitly depends on the initial temperature of the
system. I vary the assumed plasma shear viscosity to entropy density ratio 
$4 \pi \eta/{\cal S} = \{1,2,3\}$ and for zero impact parameter collisions I use central 
temperatures of $T_0 =\{520,504,494\}$ MeV, respectively.  The central temperature for 
$4 \pi \eta/{\cal S} = 1$ is fixed based on the 3+1d viscous hydrodynamical simulation
of Schenke et al which reproduces the particle spectra and elliptic flow
seen in $\sqrt{s_{NN}} = 2.76$ TeV Pb-Pb collisions \cite{Schenke:2011tv}.  The central
temperatures at $4 \pi \eta/{\cal S} = \{1,2,3\}$ were chosen in order 
to keep $dN_{\rm ch}/dy = 1400$ fixed for different assumed viscosities.  For the rapidity
dependence of the initial temperature, I use the Gaussian rapidity profile specified  in 
Ref.~\cite{Martinez:2010sd}.  Finally, I assume an 
initial anisotropy of $\xi_0=0$.  Since the current {\sc aHydro} implementation does not include transverse
dynamics, I model the transverse evolution at zero and finite impact parameter as a set of decoupled longitudinally expanding
plasmas with initial temperatures given by $T({\bf x}_\perp,b) = 
T_0 \,[N_{\rm part}({\bf x}_\perp,b)/N_{\rm part}({\bf 0},0)]^\frac{1}{3}$, where the
participant density is computed using the Glauber model with a Woods-Saxon profile and
$\sigma_{NN} = $ 62 mb.

At each transverse point I then evolve the system using {\sc aHydro} starting from $\tau_0=0.3$ fm/c, 
terminating the evolution at a final time, $\tau_f$, when the local energy density becomes less than that 
of an $N_c=3$ and $N_f=2$ ideal gas of quark and gluons with a temperature
of $T = 192$ MeV. At this temperature plasma screening effects are assumed to decrease
rapidly due to the transition to the hadronic phase and the widths of the states will become
approximately equal to their vacuum widths.


\vspace{1mm}
\noindent{\sc Formation time:}
It is also important to consider the effect of the time-dilated formation time on the
suppression of the $\Upsilon(1s)$ and $\chi_{b1}$ states~\cite{Karsch:1987uk}.  
The formation time of
a state can be estimated by the inverse of its vacuum binding energy.  Here I take  
$\tau_{\rm form}^0=\{0.2,0.4\}$ fm/c for the 
$\Upsilon(1s)$ and $\chi_{b1}$, respectively, in their local rest
frame. These formation times are consistent with the states' respective vacuum binding energies.  In the lab frame the
formation time depends
on the transverse momentum of the state via the gamma factor $\tau_{\rm form}(p_T) = \gamma 
\tau_{\rm form}^0 = E_T \tau_{\rm form}^0/M$ where $M$ is the mass of the relevant state.
For averaging over transverse momenta I assume that both states 
have a $1/E_T^4$ spectrum which is consistent with the high-$p_T$ spectrum measured by CDF
\cite{Acosta:2001gv}.

\begin{figure}[t]
\begin{center}
\includegraphics[width=\imgsize]{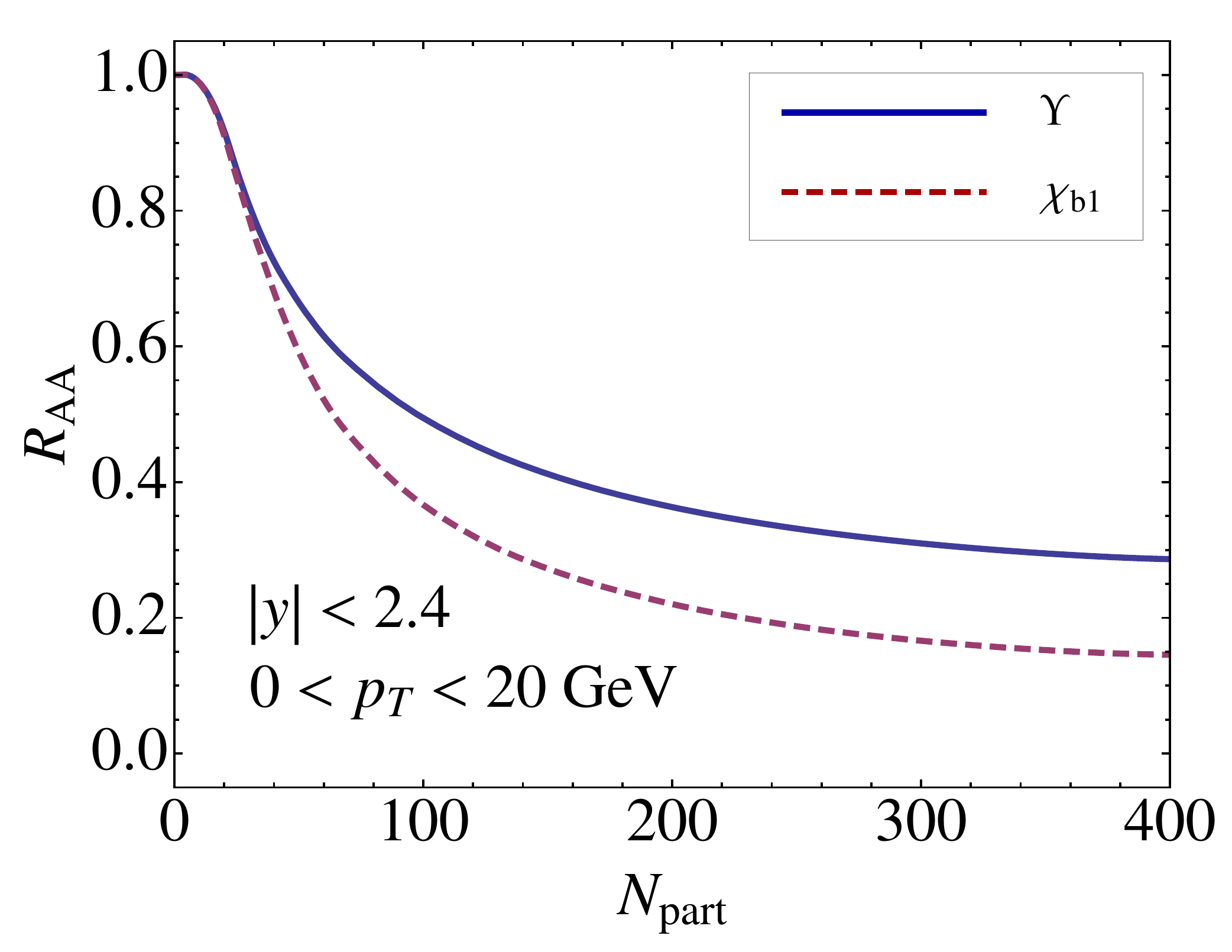}
\end{center}
\vspace{-7mm}
\caption{Rapidity- and $p_T$-averaged $R_{AA}$ for $\Upsilon(1s)$ and $\chi_{b1}$ as a function of $N_{\rm part}$ using
$4 \pi \eta/{\cal S}=1$.}
\label{fig:raasep}
\end{figure}

\begin{figure}[t]
\begin{center}
\includegraphics[width=\imgsize]{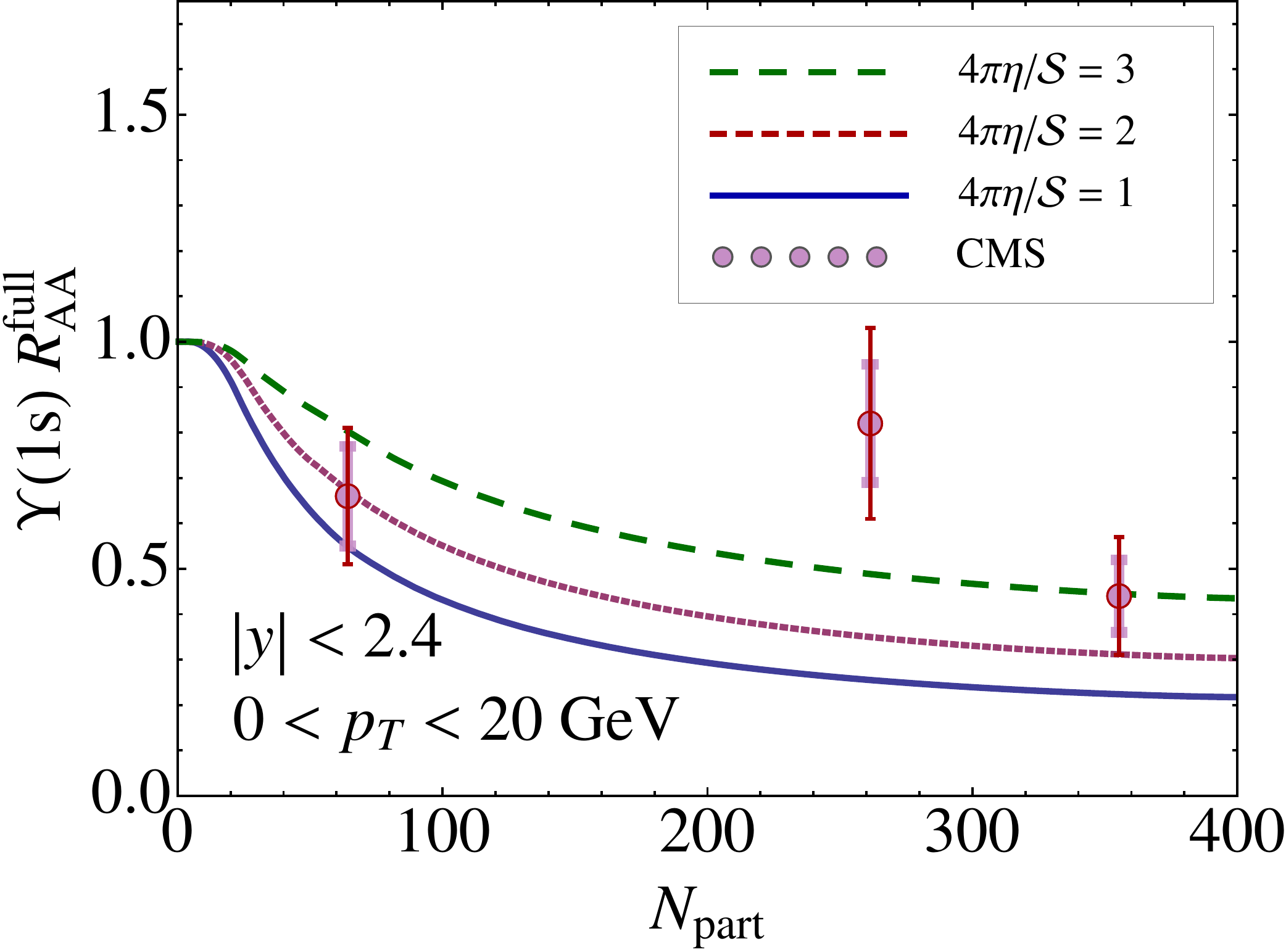}
\end{center}
\vspace{-7mm}
\caption{Rapidity- and $p_T$-averaged $R_{AA}^{\rm full}$ for the $\Upsilon(1s)$ as a function of $N_{\rm part}$
for $4 \pi \eta/{\cal S} \in \{1,2,3\}$.}
\label{fig:raanpart}
\end{figure}


\vspace{1mm}
\noindent{\sc The suppression factor:}
Having obtained the spatial and temporal evolution of the widths I can
compute the resulting nuclear suppression factor, $R_{AA}$.  Starting from the extracted time-, transverse-coordinate-,
and rapidity-dependent instantaneous decay rate, I integrate to obtain $\bar{\gamma}({\bf x}_\perp,p_T,
\varsigma,b) \equiv \Theta(\tau_f-\tau_{\rm form}(p_T)) \int_{{\rm max}(\tau_{\rm form}(p_T),\tau_0)}^{\tau_f} 
d\tau\,\Gamma_T(\tau,{\bf x}_\perp,\varsigma,b) $, in proper time \cite{Noronha:2009da} 
where $\varsigma$ is the spatial
rapidity.  From this I extract $R_{AA}$ via $R_{AA}({\bf x}_\perp,p_T,\varsigma,b) =%
\exp\!\left(-\bar{\gamma}({\bf x}_\perp,p_T,\varsigma,b) \right)$.  Finally, I average
over ${\bf x}_\perp$ taking into account the transverse dependence of the hard-particle 
production probability via 
$\langle R_{AA}(p_T,\varsigma,b) \rangle \equiv 
[\int_{{\bf x}_\perp} \! d{\bf x}_\perp \, T_{AA}({\bf x}_\perp)\,%
R_{AA}({\bf x}_\perp,p_T,\varsigma,b)]/[\int_{{\bf x}_\perp} \! d{\bf x}_\perp \,%
T_{AA}({\bf x}_\perp)]$.


\vspace{1mm}
\noindent{\sc Feed down:}
Since a certain fraction of $\Upsilon(1s)$ states produced in high energy collisions come
from the decay of excited states, when computing the total $R_{AA}$ for the $\Upsilon(1s)$
one must also consider the suppression of the excited states which decay or ``feed down'' to it.
In this work I have only computed the $R_{AA}$ for one excited state ($\chi_{b1}$), so I can
only estimate the full feed down effect.  I will estimate the full feed down effect by assuming 
that all excited states have the same $R_{AA}$ as the $\chi_{b1}$ (shown in Fig.~\ref{fig:raasep}).  
With this assumption the $\Upsilon(1s)$ $R_{AA}$ including feed down can be written as 
$R_{AA}^{\rm full} = x R_{AA}^{\rm ground\;state} + (1-x) R_{AA}^{\rm excited\;states}$
where $x$ is the percentage of $\Upsilon(1s)$ states which are produced directly.  Measurements 
of bottomonium state feed down in $\sqrt{s} = $ 1.8 TeV pp collisions at CDF \cite{Affolder:1999wm} 
with a cut $p_T^\Upsilon > $ 8.0 GeV/c find that the percentage of directly produced
$\Upsilon(1s)$ states is [50.9 $\pm$ 8.2 (stat) $\pm$ 9.0 (syst)]\%.  In all plots shown
I use $x = 0.51$.


\vspace{1mm}
\noindent{\sc Results:}
In Fig.~\ref{fig:raasep} I show the $N_{\rm part}$ dependence of $R_{AA}$ for 
$\Upsilon(1s)$ and $\chi_{b1}$.  As can be seen from this figure, despite the
fact that the initial temperature is not high enough to completely dissociate the 
$\Upsilon(1s)$, there is still a suppression due to the in-medium decay.  At these energies
we see a somewhat similar suppression pattern for the $\chi_{b1}$.  This may seem paradoxical
since the naive melting temperature for the $\chi_{b1}$ is $\sim$ 345 MeV; however, it is
important to consider the finite formation time of the $\chi_{b1}$ and the transverse
dependence of the temperature.  In practice, one finds that there are still
quite a few $\chi_{b1}$'s generated from the low temperature regions.
After averaging over the $p_T$-spectrum and geometry, the final result is the one shown 
in the Fig.~\ref{fig:raasep}.

\begin{figure}[t]
\begin{center}
\includegraphics[width=\imgsize]{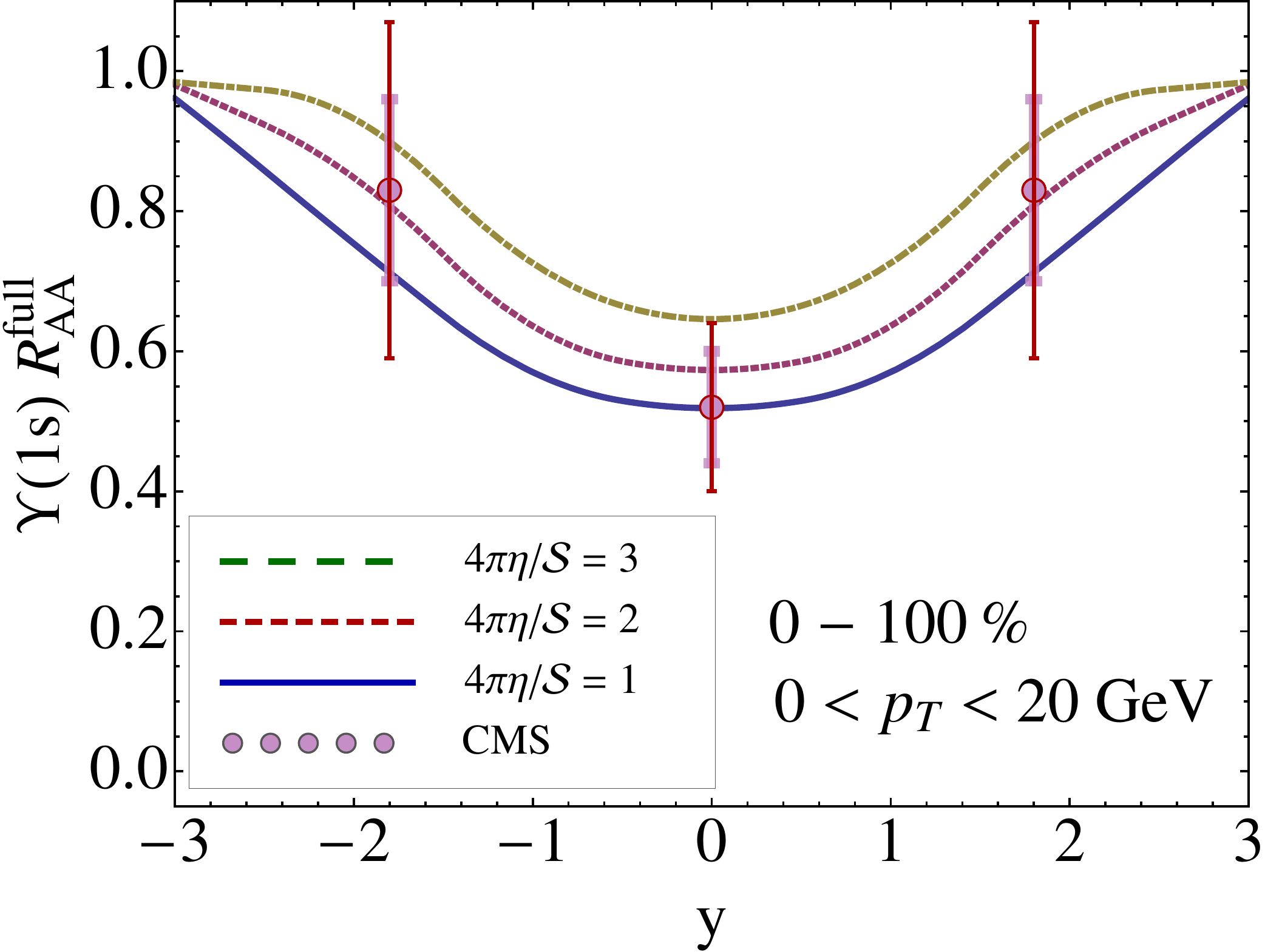}
\end{center}
\vspace{-7mm}
\caption{Centrality- and $p_T$-averaged $R_{AA}^{\rm full}$ for the $\Upsilon(1s)$ as a 
function of rapidity for $4 \pi \eta/{\cal S} \in \{1,2,3\}$.}
\label{fig:raarapeta}
\end{figure}

\begin{figure}[t]
\begin{center}
\includegraphics[width=\imgsize]{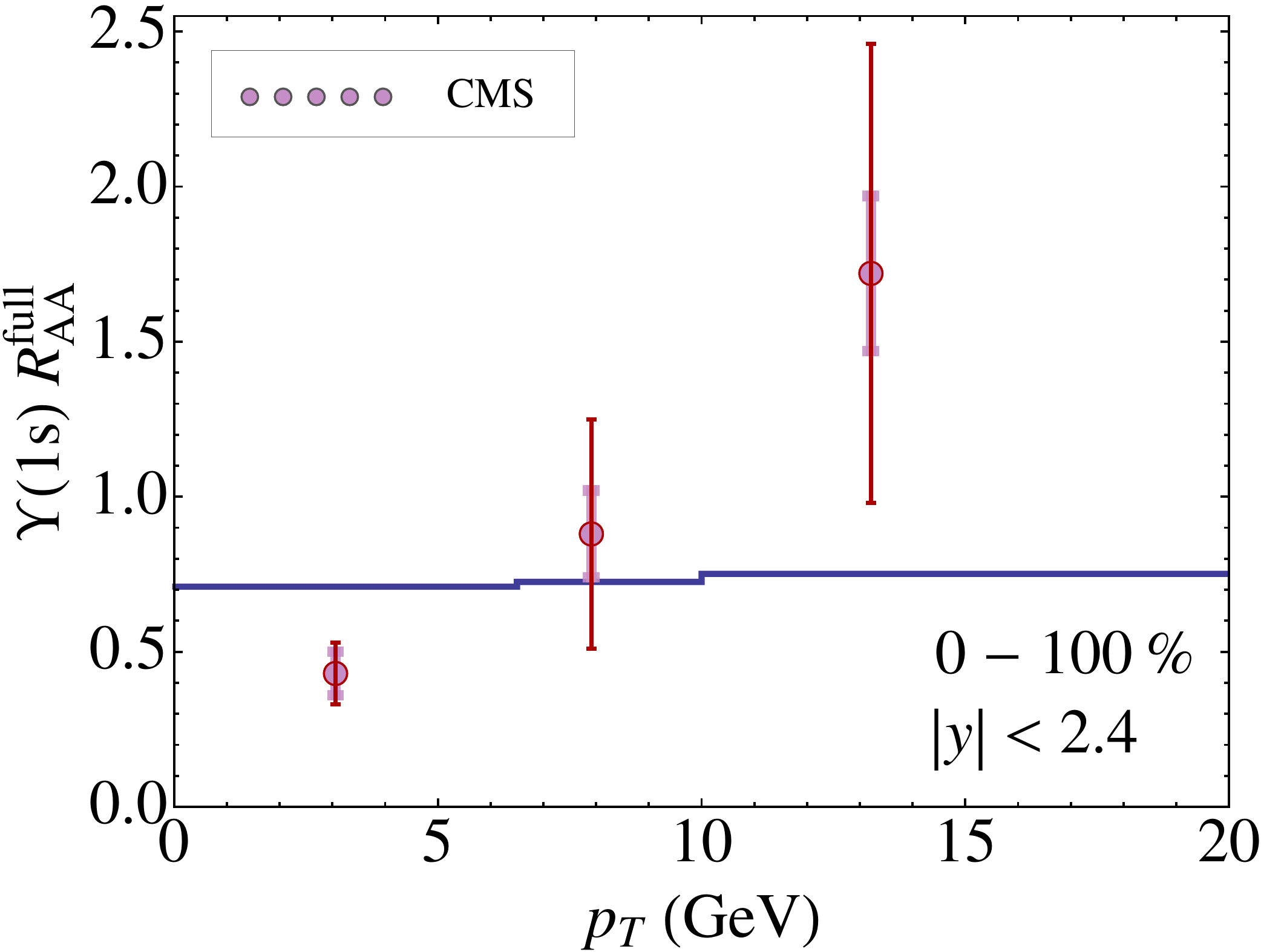}
\end{center}
\vspace{-7mm}
\caption{Centrality- and rapidity-averaged $R_{AA}^{\rm full}$ for the $\Upsilon(1s)$ as a 
function of rapidity for $4 \pi \eta/{\cal S} \in \{1,2,3\}$.}
\label{fig:raapt}
\end{figure}

In  Fig.~\ref{fig:raanpart} I show the $N_{\rm part}$ dependence of $R_{AA}^{\rm full}$
for three different values of the plasma shear viscosity, $\eta$.
The experimental data points shown are preliminary data from the CMS collaboration
\cite{HIN-10-006}.  In all cases, the statistical error is indicated by the narrow darker
(dark red online) error bar while the systematic error is indicated by the broader (purple online) shaded region.
In Fig.~\ref{fig:raarapeta} I show the rapidity dependence of the 
$\Upsilon(1s)$ $R_{AA}^{\rm full}$.  
In  Fig.~\ref{fig:raapt} I show the $p_T$ dependence of $R_{AA}^{\rm full}$
for $4 \pi \eta/{\cal S} = 1$.  In this figure I averaged over the $p_T$-bins specified by
the experiment, $0 \le p_T \le 6.5$ GeV, 6.5 GeV $ \le p_T \le 10$ GeV, and 10 GeV $\le  p_T \le 20$ GeV.

Figures.~\ref{fig:raanpart}, \ref{fig:raarapeta}, and \ref{fig:raapt} taken together demonstrate 
a reasonably good agreement between theory and experiment; however, the $p_T$ 
dependence of the theoretical result seems to be much flatter than the experimental results.  
Practically speaking though, 
based on the limited number of events, it is hard to draw firm conclusions.
Looking forward it will be important to include the suppression of the other 
relevant excited states, e.g. $\Upsilon(2s)$ and $\Upsilon(3s)$, and the effect
of transverse expansion in the dynamics.  Both of these goals represent work
in progress.


\vspace{1mm}

\noindent{\sc Acknowledgments:} 
I thank, in particular, A.~Dumitru for many useful discussions.
I also thank A.~Mocsy, P.~Petreczky, R.~Rapp, B.~Schenke, C.~Silvestre, and
 the organizers of the  BNL Summer Program ``Quarkonium 
Production in Elementary and Heavy Ion Collisions.''  Support for this project was provided 
by the Helmholtz International Center for FAIR LOEWE program and the Kavli Institute for 
Theoretical Physics Grant No. NSF PHY05-51164.

\vspace{-7mm}

\bibliography{bottomonium}

\end{document}